\pdfoutput=1

\documentclass[fleqn,twocolumn,10pt]{wlscirep}

\usepackage{macros}

\begin{document}

\title{A Formally Certified End-to-End Implementation of Shor's Factorization Algorithm}

\author[1,2]{Yuxiang Peng}
\author[1]{Kesha Hietala}
\author[3]{Runzhou Tao}
\author[1,2]{Liyi Li}
\author[4]{Robert Rand}
\author[1,2]{Michael Hicks}
\author[1,2, *]{Xiaodi Wu}

\affil[1]{Department of Computer Science, University of Maryland, College Park, USA}
\affil[2]{Joint Center for Quantum Information and Computer Science, University of Maryland, College Park, USA.}
\affil[3]{Department of Computer Science, Columbia University, USA}
\affil[4]{Department of Computer Science, University of Chicago, USA}
\affil[*]{\href{mailto:xiaodiwu@umd.edu}{xiaodiwu@umd.edu}}

\def\authorrunning{Y. Peng, K. Hietala, R. Tao, L. Li, R. Rand, M. Hicks, \& X. Wu}
\newcommand{\shortauthors}{Y. Peng, K. Hietala, R. Tao, L. Li, R. Rand, M. Hicks, \& X. Wu}

\begin{abstract}
  Quantum computing technology may soon deliver revolutionary
  improvements in algorithmic performance, but these are only useful
  if computed answers are correct. While hardware-level decoherence
  errors have garnered significant attention, a less recognized
  obstacle to correctness is that of human programming
  errors---``bugs''. Techniques familiar to most programmers from the
  classical domain for avoiding, discovering, and diagnosing bugs do not
  easily transfer, at scale, to the quantum domain because of its
  unique characteristics. To address this problem, we have been
  working to adapt \emph{formal methods} to quantum
  programming. With such methods, a programmer writes a mathematical
  specification alongside their program, and semi-automatically
  proves the program correct with respect to it. The proof's validity is automatically
  confirmed---\emph{certified}---by a ``proof assistant''. Formal
  methods have successfully yielded high-assurance
  classical software artifacts, and the underlying technology has
  produced certified proofs of major mathematical
  theorems. As a demonstration of the feasibility of applying formal
  methods to quantum programming, we present the first formally
  certified end-to-end implementation of Shor's prime factorization 
  algorithm, developed as part of a novel framework for applying the
  certified approach to general applications. By leveraging our
  framework, one can significantly reduce the effects of human errors
  and obtain a high-assurance implementation of large-scale quantum
  applications in a principled way.
\end{abstract}


\maketitle

\balance

\section{Introduction}

Leveraging the bizarre characteristics of quantum mechanics, quantum computers promise revolutionary improvements in our ability to tackle classically intractable problems, including the breaking of crypto-systems, the simulation of quantum physical systems, and the solving of various optimization and machine learning tasks.

\paragraph{Problem: Ensuring quantum programs are correct}

\begin{figure*}[!ht]
    \begin{center}
        \includegraphics[scale=0.6, trim=1cm 1.7cm 0cm 1.5cm]{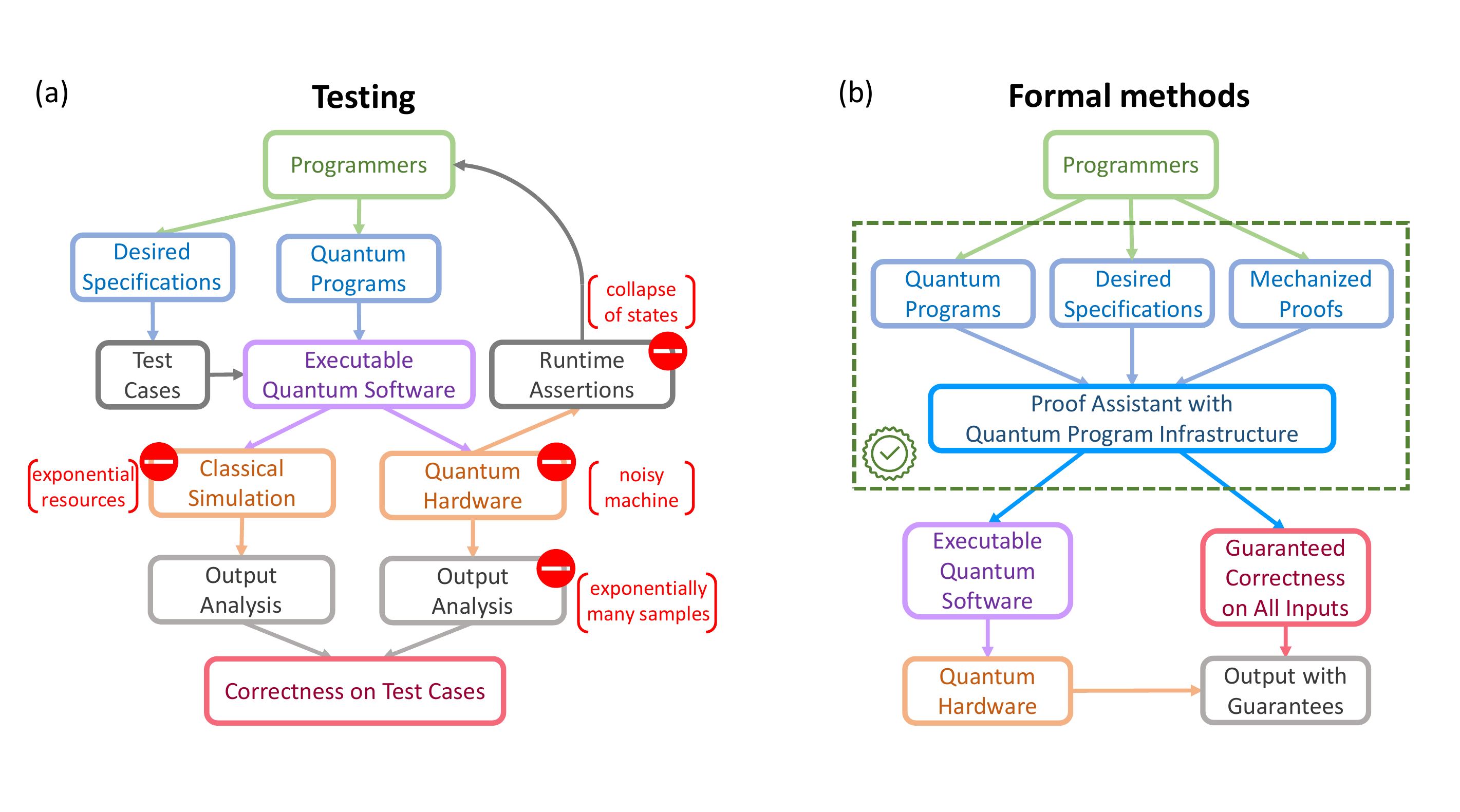}
    \caption{Comparison of developing quantum programs with testing (a) and with formal methods (b). 
    In the testing scheme, programmers will generate test cases according to the specifications of the desired quantum semantics of the target applications, and execute them on hardware for debugging purposes. 
    One approach is through runtime assertions on the intermediate program states during the execution.
    Intermediate quantum program states, however, 
    will collapse when observed for intermediate values, which implies that assertions could disturb the quantum computation itself. 
    Moreover, many quantum algorithms generate samples over an exponentially large output domain, whose statistical properties could require exponentially many samples to be verified information-theoretically. 
    Together with the fact that quantum hardware is noisy and error-prone, interpreting the readout statistics of quantum hardware for testing purposes is extremely expensive and challenging. 
    One can avoid the difficulty of working with quantum hardware by simulating quantum programs on classical machines, which, however, requires exponential resources in simulation and is not scalable at all.  
    Finally, correctness is only guaranteed on test cases in this scheme. 
    In the formal methods approach, programmers will develop quantum programs, their desired specifications, and mechanized proofs that the two correspond. All these three components---programs, specifications, and proofs---will be validated statically by the compiler of a proof assistant with built-in support to handle quantum programs. 
    Once everything passes the compiler's check, one has a certified implementation of the target quantum application, which is guaranteed to meet desired specifications on all possible inputs, even without running the program on any real machine. }  
    \label{fig:testing-vs-verification}
    \end{center}
\end{figure*}

As developments in quantum computer hardware bring this promise closer to reality, a key question to contend with is: \emph{How can we be sure that a quantum computer program, when executed, will give the right answer?} A well-recognized threat to correctness is the quantum computer hardware, which is susceptible to decoherence errors. Techniques to provide hardware-level fault tolerance are under active research~\cite{nature-fault-tolerant, RevModPhys.87.307}. A less recognized threat comes from errors---\emph{bugs}---in the program itself, as well as errors in the software that prepares a program to run on a quantum computer (compilers, linkers, etc.). In the classical domain, program bugs are commonplace and are sometimes the source of expensive and catastrophic failures or security vulnerabilities. There is reason to believe that writing correct quantum programs will be even harder, as shown in \Cref{fig:testing-vs-verification} (a).

Quantum programs that provide a performance advantage over their classical counterparts are challenging to write and understand. They often involve the use of randomized algorithms, and they leverage unfamiliar quantum-specific concepts, including superposition, entanglement, and destructive measurement. Quantum programs are also hard to test. To debug a failing test, programmers cannot easily observe (measure) an intermediate state, as the destructive nature of measurement could change the state, and the outcome. Simulating a quantum program on a classical computer can help, but is limited by such computers' ability to faithfully represent a quantum state of even modest size (which is why we must build quantum hardware). The fact that near-term quantum computers are error-prone adds another layer of difficulty.

\paragraph{Proving programs correct with formal methods}

As a potential remedy to these problems, we have been exploring how to use \emph{formal methods} (aka \emph{formal verification}) to develop quantum programs. Formal methods are processes and techniques by which one can mathematically \emph{prove} that software does what it should, for all inputs; the proved-correct artifact is referred to as \emph{formally certified}. The development of formal methods began in the 1960s when classical computers were in a state similar to quantum computers today: Computers were rare, expensive to use, and had relatively few resources, e.g., memory and processing power. Then, programmers would be expected to do proofs of their programs’ correctness by hand. Automating and confirming such proofs has, for more than 50 years now, been a grand challenge for computing research~\cite{JACM-VCompiler}. 

While early developments of formal methods led to disappointment~\cite{10.1145/359104.359106}, the last two decades have seen remarkable progress. Notable successes include the development of the seL4 microkernel \cite{sel4} and the CompCert C compiler \cite{compcert}. 
For the latter, the benefits of formal methods have been demonstrated empirically: Using sophisticated testing techniques, researchers found hundreds of bugs in the popular mainstream C compilers \texttt{gcc} and \texttt{clang}, but none in CompCert's verified core \cite{Yang2011}.
Formal methods have also been successfully deployed to prove major mathematical theorems (e.g., the Four Color theorem~\cite{gonthier2008formal}) and build computer-assisted proofs in the grand unification theory of mathematics~\cite{FM-Nature,hartnett2021}.

\paragraph{Formal methods for quantum programs}

Our key observation is that the symbolic reasoning behind the formal verification is not limited by the aforementioned
difficulties of testing directly on quantum machines or the
classical simulation of quantum machines, which lends itself
to a viable alternative to the verification of quantum programs. 
Our research has explored how formal methods can be used with quantum programs.

As shown in \Cref{fig:testing-vs-verification} (b), to develop quantum programs with formal methods we can employ a \emph{proof assistant}, which is a general-purpose tool for defining mathematical structures, and for semi-automatically mechanizing proofs of properties about those structures. The proof assistant confirms that each mechanized proof is logically correct. Using the Coq proof assistant~\cite{Coq:manual}, we defined a \emph{simple quantum intermediate representation}~\cite{hietala2021verified} (SQIR) for expressing a quantum program as a series of operations---essentially a kind of circuit---and specified those operations' mathematical meaning. Thus we can state mathematical properties about a SQIR program and prove that they always hold without needing to run that program. Then we can ask Coq to \emph{extract} the SQIR program to OpenQASM 2.0~\cite{cross2017open} to run it on specific inputs on a real machine, assured that it is correct.

Adapting formal methods developed for classical programs to work on
quantum ones are conceptually straightforward but pragmatically challenging. Consider that classical program states are (in the simplest terms) maps from addresses to bits (0 or 1); thus, a state is essentially a length-$n$ vector of booleans. Operations on states, e.g., ripple-carry adders, can be defined by boolean formulae and reasoned about symbolically.

Quantum states are much more involved: In SQIR an $n$-qubit quantum state is represented as a length-$2^n$ vector of complex numbers and the meaning of an $n$-qubit operation is represented as a $2^n \times 2^n$ matrix---applying an operation to a state is tantamount to multiplying the operation's matrix with the state's vector. Proofs over all possible inputs thus involve translating such multiplications into symbolic formulae and then reasoning about them.

Given the potentially large size of quantum states, such formulae could become quite large and difficult to reason about. To cope, we developed automated \emph{tactics} to translate symbolic states into normalized algebraic forms, making them more amenable to automated simplification. We also eschew matrix-based representations entirely when an operation can be expressed symbolically in terms of its action on basis states. With these techniques and others~\cite{hietala21proving}, we proved the correctness of key components of several quantum algorithms---Grover's search algorithm~\cite{grover} and quantum phase estimation (QPE)~\cite{shor1997polynomial}---and demonstrated advantages over competing approaches~\cite{qwire,qhlisabelle,qrhl,qbricks}.

With this promising foundation in place, several challenges remain. First, both Grover's and QPE are parameterized by \emph{oracles}, which are classical algorithmic components that must be implemented to run on quantum hardware. These must be reasoned about, too, but they can be large (many times larger than an algorithm's quantum scaffold) and can be challenging to encode for quantum processing, bug-free. Another challenge is proving the end-to-end properties of hybrid quantum/classical algorithms. These algorithms execute code on both classical and quantum computers to produce a final result. Such algorithms are likely to be common in near-term deployments in which quantum processors complement classical ones. Finally, end-to-end certified software must implement and reason about probabilistic algorithms, which are correct with a certain probability and may require multiple runs.

\paragraph{Shor's algorithm, and the benefit of formal methods}

To close these gaps, and thereby demonstrate the feasibility of the application of formal methods to quantum programming, we have produced the first fully certified version of Shor's prime factorization algorithm~\cite{shor1997polynomial}. This algorithm has been a fundamental motivation for the development of quantum computers and is at a scale and complexity not reached in prior formalization efforts. Shor's is a hybrid, probabilistic algorithm, and our certified implementation of it is complete with both classical and quantum components, including all needed oracles. 

\begin{figure*}[!ht]
    \begin{center}
        \includegraphics[scale=0.5, trim=1.5cm 14.5cm 1.5cm 5.8cm, clip]{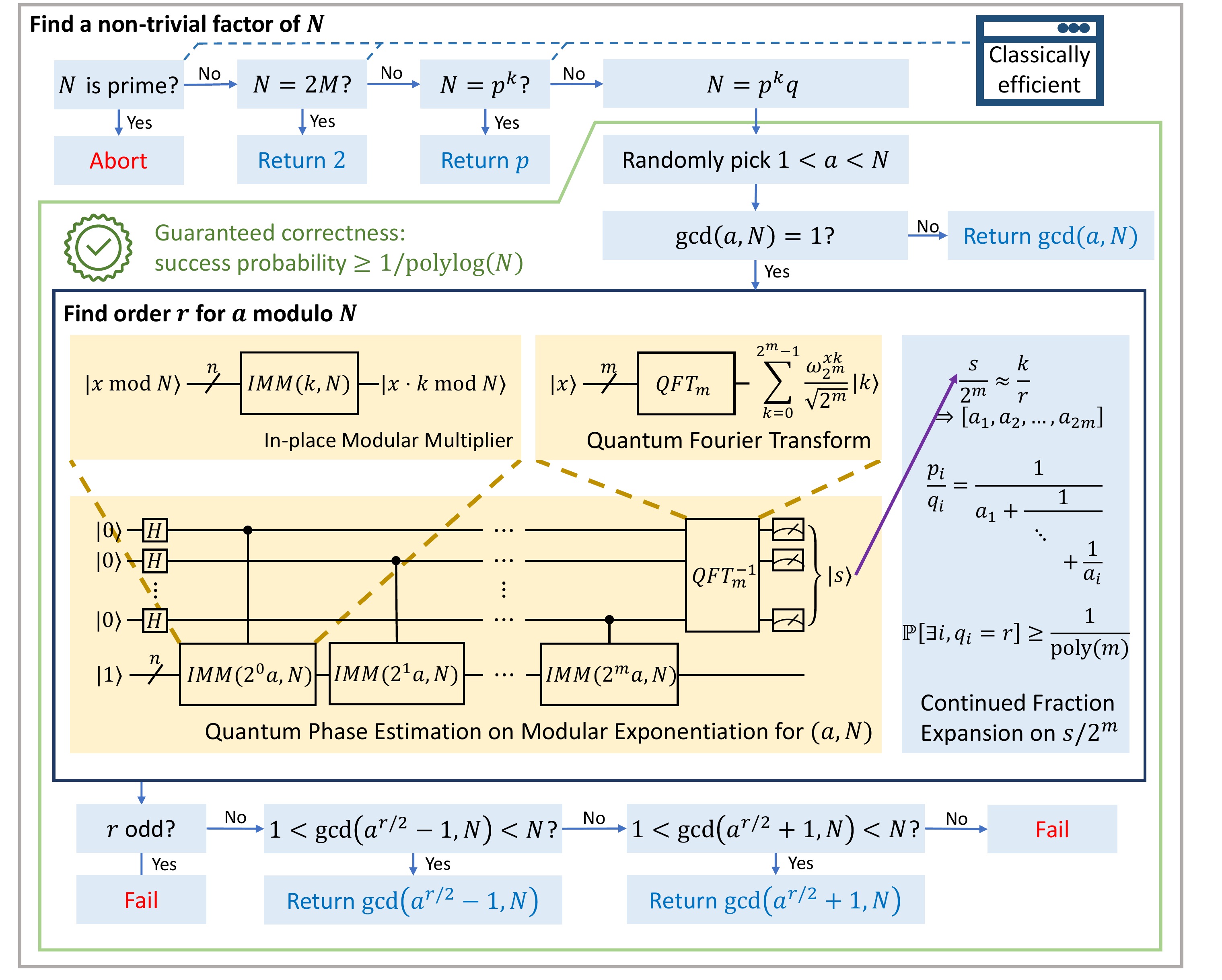}
    \caption{Overview of Shor’s factoring algorithm, which finds a non-trivial factor of integer $N$. The classical pre-processing will identify cases where $N$ is prime, even, or a prime power, which can be efficiently tested for and solved by classical algorithms. 
    Otherwise, one will proceed to the main part of Shor's algorithm (enclosed in the green frame) to solve the case where $N=p^kq$. 
    One starts with a random integer sample $a$ between $1$ and $N$. When $a$ is a co-prime of $N$, i.e., the greatest common divisor $\mathrm{gcd}(a,N)=1$, the algorithm leverages a quantum computer and classical post-processing to find the order $r$ of $a$ modulo $N$ (i.e., the smallest positive integer $r$ such that $a^r\equiv 1 (\mod N)$). 
    The quantum part of order finding involves quantum phase estimation (QPE) on modular multipliers for $(a, N)$. 
    The classical post-processing finds the continued fraction expansion (CFE) $[a_1,a_2,\cdots, a_{2m}]$ of the output $s/2^m\approx k/r$ of quantum phase estimation to recover the order $r$. 
    Further classical post-processing will rule out cases where 
    $r$ is odd before outputting the non-trivial factor. 
    To formally prove the correctness of the implementation, we first prove separately the correctness of the quantum component (i.e., QPE with in-place modular multiplier circuits for any $(a, N)$ on $n$ bits)  and the classical component (i.e., the convergence and the correctness of the CFE procedure). 
    We then integrate them to prove that with one randomly sampled $a$, the main part of Shor's algorithm, i.e., the quantum order-finding step sandwiched between the pre and post classical processing, will succeed in identifying a non-trivial factor of $N$ with probability at least 1/polylog($N$). By repeating this procedure polylog($N$) times, our certified implementation of Shor's algorithm is guaranteed to find a non-trivial factor with a success probability close to 1.}
    \label{fig:shor-overview}
    \end{center}
\end{figure*}

\section{Certified End-to-End Implementation of Shor's Prime-Factoring Algorithm}

Shor's algorithm leverages the power of quantum computing to break widely-used RSA cryptographic systems.  A recent study~\cite{Gidney2021howtofactor} suggests that with 20 million noisy qubits, it would take a few hours for Shor's algorithm to factor a 2048-bit number instead of trillions of years by modern classical computers using the best-known methods. 
As shown in \Cref{fig:shor-overview}, Shor developed a sophisticated, quantum-classical hybrid algorithm to factor a number $N$: the key quantum part---\emph{order finding}---preceded and followed by classical computation---primality testing before, and conversion of found orders to prime factors, after.
The algorithm's correctness proof critically relies on arguments about both its quantum and classical parts, and also on several number-theoretical arguments. 

While it is difficult to factor a number, it is easy to confirm a proposed factorization (the factoring problem is inside the $NP$ complexity class). One might wonder: why prove a program correct if we can always efficiently check its output? When the check shows an output is wrong, this fact does not help with computing the correct output and provides no hint about the source of the implementation error. By contrast, formal verification allows us to identify the source of the error: it's precisely the subprogram that we could not verify.

Moreover, because inputs are reasoned about \emph{symbolically}, the complexity of all-input certification can be (much) less than the complexity of single-output correctness checking. 
For example, one can symbolically verify that a quantum circuit generates a uniform distribution over $n$ bits, but
directly checking whether the output samples from a uniform distribution over $n$ bits could take as many as $2^{\Theta(n)}$ samples~\cite{test-uniform}.  
As such, with formal methods, one can certify implementation for major quantum applications, like quantum simulation which is BQP-complete~\cite{lloyd1996universal} and believed to lie outside NP.

\begin{figure*}[!htp]
    \begin{center}
        \includegraphics[width=\linewidth, trim=1.4cm 3.5cm 2.5cm 1cm, clip]{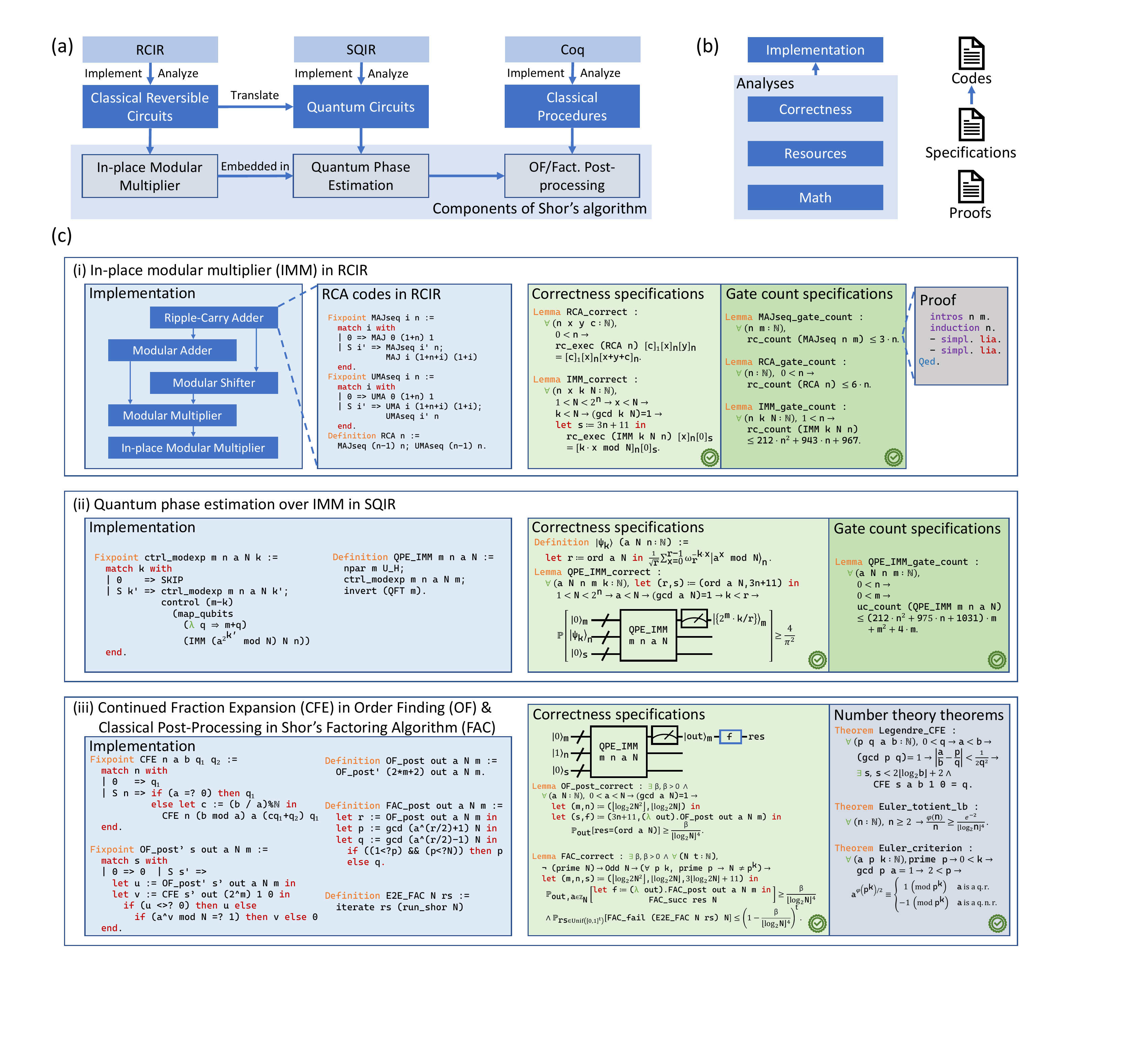}
    \caption{Technical illustration of our fully certified implementation of Shor’s algorithm. (a) The schematic framework of our implementation in Coq. \sqir is an intermediate representation of quantum circuits resembling IBM OpenQASM 2.0 but equipped with mechanized mathematical semantics in Coq. \rcir is an intermediate representation of classical reversible circuits  developed for the implementation of in-place modular multiplier (IMM) that can be translated to \sqir.    
    These three languages (Coq, \sqir, and \rcir) handle different parts of the end-to-end implementation of Shor’s algorithm as well as their correctness proof. (b) An instantiation of the formal methods scheme in Shor's implementation. Specifications of correctness and resource consumption (gate count bounds), together with their mechanized proofs (including certified math statements in number theory), are developed and validated in the Coq proof assistant.
    (c) Showcases of major components of our end-to-end implementation and corresponding proofs. Codes are adjusted for pretty-printing. (c.i) The implementation of IMM. 
    We use the example of the Ripple-Carry Adder (\coqe{RCA}) to illustrate the specifications and proofs. (c.ii) The implementation of quantum phase estimation over IMM in \sqir (\coqe{QPE\_IMM}). The correctness specification states that, under some premises, the probability of measuring the closest integer to $2^m k/r$, where $r$ is the order of $a$ modulo $N$, is larger than a positive constant $4/\pi^2$. We also certify the gate count of the implementation of \coqe{QPE\_IMM}. 
    (c.iii) The implementation of classical post-processing for order finding and factorization. Continued fraction expansion \coqe{CFE} is applied to the outcome of \coqe{QPE\_IMM} to recover the order with a certified success probability at-least 1/polylog($N$).  The success probability of factorization is also certified to be at least 1/polylog($N$), which can be boosted to 1 minus some exponentially decaying error term after repetitions. 
    These analyses critically rely on number theoretical statements like  Legendre's theorem, lower bounds for Euler's totient function, and Euler's criterion for quadratic residues, which have been proven constructively in Coq in our implementation. }
    \label{fig:implementation-overview}
    \end{center}
\end{figure*}

\paragraph{Overview of our implementation}

We carried out our work using the Coq proof assistant, using the \emph{small quantum intermediate representation} SQIR~\cite{hietala2021verified} as a basis for expressing the quantum components of Shor's algorithm. 
SQIR is a circuit-oriented quantum programming language that closely resembles IBM's OpenQASM 2.0~\cite{cross2017open} (a standard representation for quantum circuits executable on quantum machines) and is equipped with mathematical semantics using which we can reason about the properties of quantum algorithms in Coq~\cite{hietala21proving}. 
An instantiation of the scheme in \Cref{fig:testing-vs-verification} (b) for Shor's algorithm is given in \Cref{fig:implementation-overview} (b). The certified code is bounded by the green box; we proved its gate count, the likelihood of success, and correctness when successful.

The core of the algorithm is the computation of the order $r$ of $a \,\, \mathit{modulo} \,\, N$, where 
$a$ is (uniformly) randomly drawn from the numbers $1$ through $N$;
this component is bounded by the dark box in \Cref{fig:shor-overview}. The quantum component of order finding applies quantum phase estimation (QPE) to an oracle implementing an \emph{in-place modular multiplier} (IMM). 
The correctness of QPE was previously proved in SQIR with respect to an abstract oracle~\cite{hietala21proving}, but we must reason about its behavior when applied to this IMM oracle in particular. 
The oracle corresponds to pure classical reversible computation when executed coherently, leveraging none of the unique features of quantum computers, but SQIR was not able to leverage this fact to simplify the proof. 

In response, we developed the \emph{reversible circuit intermediate representation} (RCIR) in Coq to express classical functions and prove their correctness, which can be translated into SQIR as shown in \Cref{fig:implementation-overview} (a). 
RCIR helps us easily build the textbook version of IMM~\cite{ruiz2014algebraic} and prove its correctness and resource usage (\Cref{fig:implementation-overview} (c.i)). 
Integrating the QPE implementation in SQIR with the translation of IMM's implementation from RCIR to SQIR, we implement the quantum component of order-finding as well as the proof for its correctness and gate count bound (\Cref{fig:implementation-overview} (c.ii)). 
It is worth mentioning that such a proved-correct implementation of the quantum component of order finding was reported in Why3 using QBRICKS~\cite{qbricks}. 
However, the certified implementation of the classical part of order finding and the remainder of Shor's algorithm was not pursued~\cite{qbricks}. Moreover, QBRICKS' use of Why3 requires a larger trust base than Coq. 

After executing the quantum part of the algorithm, some classical code carries out continued fraction expansion (CFE) to recover the order $r$. 
Roughly speaking, the output of QPE over the IMM unitary is a close approximation of $k/r$ for a uniformly sampled $k$ from $\{0,1,\cdots, r-1\}$. 
CFE is an iterative algorithm and its efficiency 
to recover $k/r$ in terms of the number of iterations is guaranteed by Legendre's theorem which we formulated and constructively proved in Coq with respect to the CFE implementation. 
When the recovered $k$ and $r$ are co-primes, the output $r$ is the correct order. The algorithm is probabilistic, and the probability that co-prime $k$ and $r$ are output is lower bounded by the size of $\mathbb{Z}_r$ which consists of all positive integers that are smaller than $r$ and coprime to it. 
The size of $\mathbb{Z}_r$ is the definition of the famous Euler's totient function $\varphi(r)$, which we proved is at least $e^{-2}/\lfloor\log(r)\rfloor^4$ 
in Coq based on the formalization of Euler's product formula and Euler's theorem by de Rauglaudre\cite{roglo2020euler}.
By integrating the proofs for both quantum and classical components, we show that our implementation of the entire hybrid order-finding procedure will identify the correct order $r$ for any $a$ given that $\mathrm{gcd}(a,N)=1$ with probability at least 
$4e^{-2}/\pi^2\lfloor \log_2(N)\rfloor^4$ (\Cref{fig:implementation-overview} (c.iii)).

With the properties and correctness of order finding established, we can prove the success probability of the algorithm overall. In particular, we aim to establish that the order finding procedure combined with the classical post-processing will output a non-trivial factor with a success probability of at least $2e^{-2}/\pi^2\lfloor \log_2(N)\rfloor^4$, which is half of the success probability of order finding. 
In other words, we prove that for at least a half of the integers $a$ between $1$ and $N$, the order $r$ will be even and either $\mathrm{gcd}(a^{r/2}+1, N)$ or $\mathrm{gcd}(a^{r/2}-1,N)$ will be a non-trivial factor of $N$. 
Shor's original proof~\cite{shor1997polynomial} of this property made use of the existence of the group generator of $\mathbb{Z}_{p^k}$, also known as primitive roots, for odd prime $p$. However, the known proof of the existence of primitive roots is non-constructive~\cite{hardy75} meaning that it makes use of axioms like the law of the excluded middle, whereas one needs to provide constructive proofs~\cite{Bauer2010AcceptingConstructiveMath} in Coq and other proof assistants. 

To address this problem, we provide a new, constructive proof of the desired fact without using primitive roots. 
Precisely, we make use of the quadratic residues in modulus $p^k$ and connect whether a randomly chosen $a$ leads to a non-trivial factor to the number of quadratic residues and non-residues in modulus $p^k$. 
The counting of the latter is established based on Euler's criterion for distinguishing between quadratic residues and non-residues modulo $p^k$ which we have constructively proved in Coq. 

Putting it all together, we have proved that our implementation of Shor's algorithm successfully outputs a non-trivial factor with a probability of at least $2e^{-2}/\pi^2\lfloor \log_2(N)\rfloor^4$ for one random sample of $a$. 
Furthermore, we also prove in Coq that its failure probability of $t$ repetitions is upper bounded by $(1-2e^{-2}/\pi^2\lfloor \log_2(N)\rfloor^4)^t$, 
which boosts the success probability of our implementation arbitrarily close to 1 after $O(\log^4(N))$ repetitions. 

We also certify that the gate count in our implementation of Shor's algorithm using OpenQASM's gate set is upper bounded by $(212n^2+975n+1031)m+4m+m^2$ in Coq, 
where $n$ refers to the number of bits representing $N$ and $m$ the number of bits in QPE output. 
Note further $m, n=O(\log N)$, which leads to an $O(\log^3 N)$ overall asymptotic complexity that matches the original paper.  

\begin{figure*}[!htp]
    \begin{center}
        \includegraphics[scale=0.75, trim=5.5cm 13.5cm 5cm 11.3cm, clip]{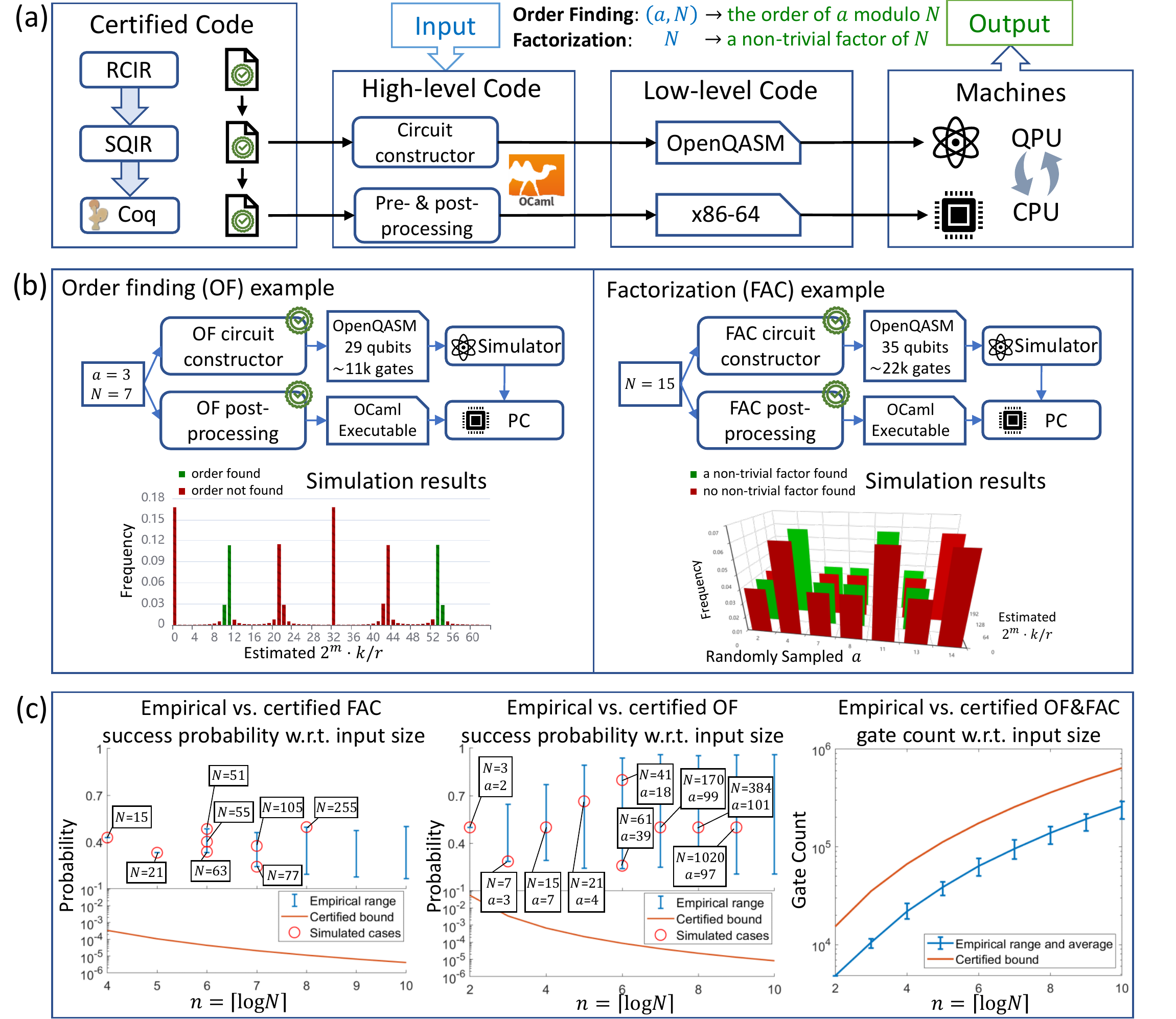}
    \caption{End-to-End Execution of Shor's algorithm. (a) A schematic illustration of the end-to-end quantum-classical hybrid execution.  Programmers write programs, specifications, and proofs in Coq, where Coq programs are extracted to OCaml for practical execution. Given an input parameter $a, N$ for order finding (or $N$ for factorization), the extracted OCaml program generates an OpenQASM file for a quantum processing unit and an executable for a classical machine to pre and post classical processing. (b) Examples of end-to-end executions of order finding (OF) and factorization (FAC). The left example finds the order for $a$=3 and $N$=7. The generated OpenQASM file uses 29 qubits and contains around 11k gates. We employed JKQ DDSIM \cite{ddsim} to simulate the circuit for 100k shots, and the frequency distribution is presented. The trials with post-processing leading to the correct order  $r$=6 are marked green. The empirical success probability is 28.40\%, whereas the proved success probability lower bound is 0.34\%. The right example shows the procedure factorizing $N$=15. For each randomly picked $a$, the generated OpenQASM file uses 35 qubits and contains around 22k gates, which are simulated by JKD DDSIM with the outcome frequency presented in the figure. 
    The cases leading to a non-trivial factor are marked green. The empirical success probability is 43.77\%, whereas the proved success probability lower bound is 0.17\%. 
    (c) Empirical statistics of the gate count and success probability of order finding and factorization for every valid input $N$ with respect to input size $n$ from 2 to 10 bits. We draw the bounds certified in Coq as red curves.
    Whenever the simulation is possible with DDSIM, we draw the empirical bounds as red circles. 
    Otherwise, we compute the corresponding bounds using  analytical formulas with concrete inputs. These bounds are drawn as blue intervals called empirical ranges (i.e., minimal to maximal success probability) for each input size. 
    }
    \label{fig:experiments}
    \end{center}
\end{figure*}

\section{Executing Shor's algorithm}

Having completed our certified-in-Coq implementation of Shor's algorithm, we \emph{extract} the program---both classical and quantum parts---to code we can execute. Extraction is simply a lightweight translation from Coq's native language to Objective Caml (OCaml), a similar but an executable alternative~\cite{coqextraction} which runs on a classical computer.
The quantum part of Shor's algorithm is extracted to OCaml code that, when executed, generates the desired quantum circuits in OpenQASM 2.0 for the given input parameters; this circuit will be executed on a quantum computer. 
The classical pre- and post-processing codes extract directly to OCaml.
A schematic illustration of this end-to-end quantum-classical hybrid execution is given in \Cref{fig:experiments} (a) for both order finding and factorization. 

In principle, the generated Shor's factorization circuits could be executed on any quantum machine. 
However, for small instances, as we elaborate on later, the size of these quantum circuits is still challenging for existing quantum machines to execute. 
Instead, we use a classical simulator called DDSIM~\cite{ddsim} to execute these quantum circuits, which necessarily limits the scale of our empirical study. 

It is worth mentioning that experimental demonstration of Shor's algorithm already exists for small instances like $N=$15\cite{vandersypen2001experimental, lu2007demonstration, lanyon2007experimental, lucero2012computing, monz2016realization} or 21~\cite{martin2012experimental}, which uses around 5 qubits and 20 gates.
These experimental demonstrations are possible because they leverage quantum circuits that are specially designed for fixed inputs but cannot extend to work for general ones. 
In our implementation, an efficient circuit constructor will generate the desired quantum circuit given any input. Even for small instances (order finding with input $(a=3, N=7)$ and factorization with $N=15$), the generated quantum circuits would require around 30 qubits and over 10k gates, whose details of the simulator-based execution are shown in \Cref{fig:experiments} (b). 

In \Cref{fig:experiments} (c), we conduct a more comprehensive empirical study on the gate count and 
success probability of order finding and factorization instances with input size ($\log(N)$) from 2 to 10 bits, i.e., $N\leq 1024$. 
Red circles refer to instances (i.e. a few specific $N$s) that can be simulated by DDSIM. The empirical success probability for other $N$s up to 1024 
are calculated directly using formulas in Shor's original analysis with specific inputs, whereas our certified bounds are loose in the sense that they only hold for each input size. 
These empirical values are displayed in a blue interval called the empirical range per input size.  
It is observed that 
(1) certified bounds hold for all instances 
(2) empirical bounds are considerably better than certified ones for studied instances.  
The latter is likely due to the non-optimality of our proofs in Coq and the fact that we only investigated small-size instances. 

\section{Conclusions}

The nature of quantum computing makes programming, testing, and debugging quantum programs difficult, and this difficulty is exacerbated by the error-prone nature of quantum hardware. As a long-term remedy to this problem, we propose to use formal methods to mathematically certify that quantum programs do what they are meant to. To this end, we have leveraged prior formal methods work for classical programs, and extended it to work on quantum programs. As a showcase of the feasibility of our proposal, we have developed the first formally certified end-to-end implementation of Shor's prime factorization algorithm. 

The complexity of software engineering of quantum applications would grow significantly with the development 
of quantum machines in the near future. 
We believe that our proposal is a principled approach to mitigating human errors in this critical domain and achieving high assurance for important quantum applications.

\section*{Acknowledgement}
We thank Andrew Childs, Steven Girvin, Liang Jiang, and Peter Shor for helpful feedback on the manuscript. 
This material is based upon work supported by the Air Force Office of Scientific Research under award  number FA95502110051, the U.S. Department of Energy, Office of Science, Office of
Advanced Scientific Computing Research, Quantum Testbed Pathfinder Program under Award
Number DE-SC0019040, and the U.S. National Science Foundation grant CCF-1942837 (CAREER). Any opinions, findings, conclusions, or recommendations expressed in this material are those of the  author(s) and do not necessarily reflect the views of these agencies.

\bibliography{refs,qrefs}

\newpage
\newpage 
\appendix

\begin{center}
 \Large \bf Supplementary Materials
\end{center}

All codes in the implementation are available at \url{https://github.com/inQWIRE/SQIR/tree/main/examples/shor}.
The entire implementation includes approximately 14k lines of code. 

\section{Preliminaries in Formal Methods}

We assume a background in quantum computing. 

\subsection{Proof Assistants}

A \emph{proof assistant} is a software tool for formalizing mathematical definitions and stating and proving properties about them. A proof assistant may produce proofs automatically or assist a human in doing so, interactively. Either way, the proof assistant confirms that a proof is correct by employing a \emph{proof verifier}. Since a proof's correctness relies on the verifier being correct, a verifier should be small and simple and the logical rules it checks should be consistent (which is usually proved meta-theoretically).

Most modern proof assistants implement proof verification by leveraging the \emph{Curry-Howard correspondence}, which embodies a surprising and powerful analogy between formal logic and programming language type systems~\cite{curry1934functionality, howard1980formulae}. In particular, logical propositions are analogous to programming language types, and proofs are analogous to programs. As an example, the logical implication in proof behaves like a function in programs: Given a proof (program expression $a$) of proposition (type) $A$, and a proof that $A$ implies $B$ (a function $f$ of type $A \rightarrow B$), we can prove the proposition $B$ (produce a program expression of type $B$, i.e., via
the expression $f(a)$). We can thus represent a proof of a logical formula as a typed expression whose type corresponds to the formula. As a result, proof verification is tantamount to (and implemented as) program type checking.

Machine-aided proofs date back to the Automath project by de Bruijn \cite{de1970mathematical}, which was the first practical system exploiting the Curry-Howard correspondence. Inspired by Automath, interactive theorem provers (ITPs) emerged. Most modern proof assistants are ITPs. Milner proposed Stanford LCF \cite{milner1972implementation}, introducing \emph{proof tactics}, which allow users to specify particular automated proof search procedures when constructing a proof. A tactic reduces the current proof goal to a list of new subgoals. The process of producing a machine-aided proof  is to sequentially apply a list of tactics to transform a proof goal into predefined axioms. Users have direct access to the intermediate subgoals to decide which tactic to apply. 

While ITPs were originally developed to formalize mathematics, the use of the Curry-Howard correspondence makes it straightforward to also support writing proved-correct, i.e., \emph{verified}, computer programs. These programs can be \emph{extracted} into runnable code from the notation used to formalize them in the proof assistant.

Modern ITPs are based on different variants of type theories.
The ITP employed in this project, Coq \cite{coquand85higher}, is based on the Calculus of Inductive Constructions \cite{cic}. Coq features propositions as types, higher order logic, dependent types, and reflections. A variety of proof tactics are included in Coq, like induction. These features have made Coq widely used by the formal methods community.

Coq is a particularly exciting tool that has been used both to verify complex programs and to prove hard mathematical theorems. The archetype of a verified program is the CompCert compiler \cite{compcert}. CompCert compiles code written in the widely used C programming language to instruction sets for ARM, x86, and other computer architectures. Importantly, CompCert's design precisely reflects the intended program behavior---the \emph{semantics}---given in the C99 specification, and all of its optimizations are guaranteed to preserve that behavior. Coq has also been used to verify proofs of the notoriously hard-to-check Four Color Theorem, as well as the Feit–Thompson (or odd order) theorem. Coq's dual uses for both programming and mathematics make it an ideal tool for verifying quantum algorithms.

Coq isn't the only ITP with a number of success stories. The F$^*$ language is being used to certify a significant number of internet security protocols, including Transport Layer Security (TLS) \cite{everest} and the High Assurance Cryptographic Library, HACL$^*$ \cite{haclstar}, which has been integrated into the Firefox web browser. Isabelle/HOL was used to verify the seL4  operating system kernel \cite{sel4}. The Lean proof assistant (also based on the Calculus of Inductive Constructions) has been used to verify essentially the entire undergraduate mathematics curriculum and large parts of a graduate curriculum \cite{leanmath}. Indeed, Lean has reached the point where it can verify cutting-edge proofs, including a core theorem in Peter Scholze's theory of condensed mathematics, first proven in 2019~\cite{castelvecchi2021, hartnett2021}. 
Our approach to certifying quantum programs could be implemented using these other tools as well.

\subsection{\sqir}

To facilitate proofs about quantum programs, we developed the \emph{small quantum intermediate representation} (\sqir)\cite{hietala2021verified, hietala21proving}, a circuit-oriented programming language \emph{embedded} in Coq, which means that a \sqir program is defined as a Coq data structure specified using a special syntax, and the semantics of a \sqir program is defined as a Coq function over that data structure (details below). We construct quantum circuits using \sqir, and then state and prove specifications using our Coq libraries for reasoning about quantum programs. \sqir programs can be extracted to OpenQASM 2.0 \cite{cross2017open}, a standard representation for quantum circuits executable on quantum machines.

A \sqir program is a sequence of gates applied to natural number arguments, referring to names (labels) of qubits in a global register. Using predefined gates \coqe{SKIP} (no-op), \coqe{H} (Hadamard), and \coqe{CNOT} (controlled \emph{not}) in \sqir, a circuit that generates the Greenberger–Horne–Zeilinger (GHZ) state with three qubits in Coq is defined by
\begin{coq}
Definition GHZ3 : ucom base 3 := H 0; CNOT 0 1; CNOT 0 2.
\end{coq}
The type \coqe{ucom base 3} says that the resulting circuit is a unitary program that uses our base gate set and three qubits. Inside this circuit, three gates are sequentially applied to the qubits. More generally, we could write a Coq function that produces a GHZ state generation circuit: Given a parameter $n$, function \coqe{GHZ} produces the $n$-qubit GHZ circuit. 
\begin{coq}
Fixpoint GHZ (n : $\mathbb{N}$) : ucom base n :=
  match n with
  | 0 => SKIP
  | 1 => H 0
  | S (S n') => GHZ (S n'); CNOT n' (S n')      
  end.
\end{coq}
These codes define a recursive prograom \texttt{GHZ} on one natural number input \texttt{n} through the use of \texttt{match} statement. Specifically, \texttt{match} statement returns \texttt{SKIP} when \texttt{n=0}, \texttt{H 0} when \texttt{n=1}, and recursively calls on itself for \texttt{n-1} otherwise.
One can observe that \coqe{GHZ  3} (calling \coqe{GHZ} with argument $3$) will produce the same \sqir circuit as definition \coqe{GHZ3}, above.

The function \coqe{uc_eval} defines the semantics of a \sqir program, essentially by converting it to a unitary matrix of complex numbers. This matrix is expressed using axiomatized reals from the Coq Standard Library \cite{coqreals}, complex numbers from Coquelicot \cite{coquelicot}, and the complex matrix library from \qwire~\cite{paykin2017qwire}. Using \coqe{uc_eval}, we can state properties about the behavior of a circuit. For example, the specification for \coqe{GHZ} says that it produces the mathematical $GHZ$ state when applied to the all-zero input.
\begin{coq}
Theorem GHZ_correct : forall n : $\mathbb{N}$, 0 < n ->
    uc_eval (GHZ n) $\times$ $|0\rangle^{\otimes \texttt{n}}$ = $\frac{1}{\sqrt{2}}$ * $|0\rangle^{\otimes \texttt{n}}$ + $\frac{1}{\sqrt{2}}$ * $|1\rangle^{\otimes \texttt{n}}$.
\end{coq}
This theorem can be proved in Coq by induction on $n$.

To date, \sqir has been used to implement and verify a number of quantum algorithms \cite{hietala21proving}, including quantum teleportation, GHZ state preparation, the Deutsch-Jozsa algorithm, Simon’s algorithm, the quantum Fourier transform (QFT), Grover’s algorithm, and quantum phase estimation (QPE). QPE is a key component of Shor’s prime factoring algorithm (described in the next section), which finds the eigenvalue of a quantum program's eigenstates. 

Using \sqir, we define QPE as follows:
\begin{coq}
Fixpoint controlled_powers {n} f k kmax :=
  match k with
  | 0    => SKIP
  | 1    => control (kmax-1) (f O)
  | S k' => controlled_powers f k' kmax ;
            control (kmax-k'-1) (f k')
  end.
Definition QPE k n (f : nat -> base_ucom n) :=
  let f' := (fun x => map_qubits (fun q => k+q) (f x)) in
  npar k U_H ;
  controlled_powers f' k k ; 
  invert (QFT k).
\end{coq}
\coqe{QPE} takes as input the precision \texttt{k} of the resulting estimate, the number \coqe{n} of qubits used in the input program, and a circuit family \coqe{f}.
\coqe{QPE} includes three parts: (1) \coqe{k} parallel applications of Hadamard gates; (2) exponentiation of the target unitary; (3) an inverse QFT procedure. (1) and (3) are implemented by recursive calls in \sqir. Our implementation of (2) inputs a mapping from natural numbers representing which qubit is the control, to circuits implementing repetitions of the target unitary, since normally the exponentiation is decomposed into letting the $x$-th bit control $2^x$ repetition of the target unitary. Then \coqe{controlled_powers} recursively calls itself, in order to map the circuit family on the first \coqe{n} qubits to the exponentiation circuit. In Shor's algorithm, (2) is efficiently implemented by applying controlled in-place modular multiplications with pre-calculated multipliers. The correctness of \coqe{QPE} is also elaborated\cite{hietala21proving}. 

\section{Shor's Algorithm and Its Implementation}

Shor's factorization algorithm consists of two parts. The first employs a hybrid classical-quantum algorithm to solve the order finding problem; the second reduces factorization to order finding. In this section, we present an overview of Shor's algorithm (see \Cref{fig:shor-overview} for a summary). In next sections, we discuss details about our implementation (see \Cref{fig:implementation-overview}) and certified correctness properties.

\subsection{A Hybrid Algorithm for Order Finding}
\label{app:imp:of}

The multiplicative order of $a$ modulo $N$, represented by $\text{ord}(a, N)$, is the least integer $r$ larger than $1$ such that $a^r\equiv 1 \pmod{N}.$ Calculating $\text{ord}(a, N)$ is hard for classical computers, but can be efficiently solved with a quantum computer, for which Shor proposed a hybrid classical-quantum algorithm \cite{shor1997polynomial}. This algorithm has three major components: (1) in-place modular multiplication on a quantum computer; (2) quantum phase estimation; (3) continued fraction expansion on a classical computer.

\paragraph{In-place Modular Multiplication}
An in-place modular multiplication operator $IMM(a, N)$ on $n$ working qubits and $s$ ancillary qubits satisfies the following property:
\begin{align*}
    \forall x<N,~~ IMM(a, N)|x\rangle_n|0\rangle_s=|(a\cdot x) \text{ mod } N\rangle_n|0\rangle_s,
\end{align*}
where $0<N<2^{n-1}.$ It is required that $a$ and $N$ are co-prime, otherwise the operator is irreversible. This requirement implies the existence of a multiplicative inverse $a^{-1}$ modulo $N$ such that $a\cdot a^{-1}\equiv 1 \pmod{N}$. 

\paragraph{Quantum Phase Estimation}
Given a subroutine $U$ and an eigenvector $\ket{\psi}$ with eigenvalue $e^{i\theta},$ quantum phase estimation (QPE) finds the closest integer to $\frac{\theta}{2\pi}2^m$ with high success probability, where $m$ is a predefined precision parameter.

Shor's algorithm picks a random $a$ from $[1, N)$ first, and applies QPE on $IMM(a, N)$ on input state $\ket{0}_m\ket{1}_n\ket{0}_s$ where $m=\lfloor \log_2 2N^2 \rfloor, n=\lfloor \log_2 2N \rfloor$ and $s$ is the number of ancillary qubits used in $IMM(a, N)$. Then a computational basis measurement is applied on the first $m$ qubits, generating an output integer $0\leq \tt{out}<2^m.$ The distribution of the output has $\text{ord}(a, N)$ peaks, and these peaks are almost equally spaced. We can extract the order by the following procedure.

\paragraph{Continued Fraction Expansion}
The post-processing of Shor's algorithm invokes the continued fraction expansion (CFE) algorithm. A $k$-level continued fraction is defined recursively by
\begin{align*}
    \langle \rangle&=0, \\
    \langle a_1, a_2, ..., a_k \rangle&=\frac{1}{a_1+\langle a_2, a_3, ..., a_k\rangle}.
\end{align*}
$k$-step CFE finds a $k$-level continued fraction to approximate a given real number. For a rational number $0\leq \frac{a}{b}<1$, the first term of the expansion is $\lfloor \frac{b}{a} \rfloor$ if $a\neq 0$, and we recursively expand $\frac{b\text{ mod }a}{a}$ for at most $k$ times to get an approximation of $\frac{a}{b}$ by a $k$-level continued fraction. In Coq, the CFE algorithm is implemented as 
\begin{coq}
Fixpoint CFE_ite (k a b $\tt{p}_1~ \tt{q}_1~ \tt{p}_2~ \tt{q}_2$ : $\mathbb{N}$) : $\mathbb{N}\times\mathbb{N}$ :=
  match k with
  | 0 => ($\tt{p}_1$, $\tt{q}_1$)
  | S k' => if a = 0 then ($\tt{p}_1$, $\tt{q}_1$)
            else let (c, d) := ($\lfloor\frac{\tt{b}}{\tt{a}}\rfloor,$ b mod a) in
            CF_ite k' d a $(\tt{c}\cdot \tt{p}_1+\tt{p}_2)$ $(\tt{c}\cdot \tt{q}_1+\tt{q}_2)$ $\tt{p}_1$ $\tt{q}_1$
  end.
Definition CFE k a b := snd (CF_ite (k+1) a b 0 1 1 0).
\end{coq}
Function \coqe{CFE_ite} takes in the number of iterations $k$, target fraction \texttt{a/b}, the fraction from the $(k-1)$-step expansion, and the $(k-2)$-step expansion. Function \coqe{CFE k a b} represents the denominator in the simplified fraction equal to the $k$-level continued fraction that is the closest to $\frac{a}{b}$. 

The post-processing of Shor's algorithm expands $\frac{\tt{out}}{2^{\tt{m}}}$ using CFE, where \coqe{out} is the measurement result and \coqe{m} is the precision for QPE defined above. It finds the minimal step $\tt{k}$ such that $a^{\tt{CFE k out }2^{\tt{m}}} \equiv 1\pmod{N}$ and $k\leq 2m+1.$ With probability no less than $1/\text{polylog}(N)$, there exists $k$ such that $\tt{CFE k out }2^{\tt{m}}$ is the multiplicative order of $a$ modulo $N$. We can repeat the QPE and post-processing for $\text{polylog}(N)$ times. Then the probability that the order exists in one of the results can be arbitrarily close to $1$. The minimal valid post-processing result is highly likely to be the order. 

\subsection{Reduction from Factorization to Order Finding}

To completely factorize composite number $N$, we only need to find one non-trivial factor of $N$ (i.e., a factor that is not 1 nor $N$). If a non-trivial factor $d$ of $N$ can be found, we can recursively solve the problem by factorizing $d$ and $\frac{N}{d}$ separately. Because there are at most $\log_2(N)$ prime factors of $N$, this procedure repeats for at most $\text{polylog}(N)$ times. 
A classical computer can efficiently find a non-trivial factor in the case where $N$ is even or $N=p^k$ for prime $p$. 
However, Shor's algorithm is the only known (classical or quantum) algorithm to efficiently factor numbers for which neither of these is true.

Shor's algorithm randomly picks an integer $1\leq a<N$. If the greatest common divisor $\text{gcd}(a, N)$ of $a$ and $N$ is a non-trivial factor of $N$, then we are done. Otherwise we invoke the hybrid order finding procedure to find $\text{ord}(a, N)$. With probability no less than one half, one of $\text{gcd}\left(a^{\lfloor\frac{\text{ord}(a, N)}{2}\rfloor}\pm 1, N\right)$ is a non-trivial factor of $N$. Note that $\text{gcd}\left(a^{\lfloor\frac{\text{ord}(a, N)}{2}\rfloor}\pm1, N\right)$ can be efficiently computed by a classical computer\cite{knuth1997v1}. By repeating the random selection of $a$ and the above procedure for constant times, the success probability to find a non-trivial factor of $N$ is close to $1$.

\subsection{Implementation of Modular Multiplication}
\label{sec:imm}

One of the pivoting components of Shor's order finding procedure is a quantum circuit for in-place modular multiplication (IMM). 
We initially tried to define this operation in \sqir but found that for purely classical operations (that take basis states to basis states), \sqir's general quantum semantics makes proofs unnecessarily complicated.
In response, we developed the \emph{reversible circuit intermediate representation} (\rcir) to express classical functions and prove their correctness.
\rcir programs can be translated into \sqir, and we prove this translation correct.

\paragraph{\rcir}
\rcir~contains a universal set of constructs on classical bits labeled by natural numbers. The syntax is:
\[R ~:=~ \texttt{skip} ~~|~~ \texttt{X} ~n ~~|~~ \texttt{ctrl} ~n ~R ~~|~~ \texttt{swap} ~m ~n ~~|~~ R_1; ~R_2.\]
Here \texttt{skip} is a unit operation with no effect, $\texttt{X}~n$ flips the $n$-th bit, $\texttt{ctrl}~n~R$ executes subprogram $R$ if the $n$-th bit is $1$ and otherwise has no effect, $\texttt{swap}~m~n$ swaps the $m$-th and $n$-th bits, and $R_1; R_2$ executes subprograms $R_1$ and $R_2$ sequentially. We remark that \texttt{swap} is not necessary for the expressiveness of the language, since it can be decomposed into a sequence of three \texttt{ctrl} and \texttt{X} operations. 
We include it here to facilitate \texttt{swap}-specific optimizations of the circuit.

As an example, we show the \rcir code for the MAJ (majority) operation \cite{cuccaro2004new}, which is an essential component of the ripple-carry adder.
\begin{coq}
Definition MAJ a b c :=
  ctrl c (X b) ; ctrl c (X a) ; ctrl a (ctrl b (X c)).
\end{coq}
It takes in three bits labeled by $a, b, c$ whose initial values are $v_a, v_b, v_c$ correspondingly, and stores $v_a \texttt{ xor } v_c$ in $a$, $v_b \texttt{ xor } v_c$ in $b$, and $MAJ(v_a, v_b, v_c)$ in $c$. Here $MAJ(v_a, v_b, v_c)$ is the majority of $v_a, v_b$ and $v_c,$ the value that appears at least twice.

To reverse a program written in this syntax, we define a reverse operator by $\tt{skip}^{\tt{rev}}=\tt{skip}$, $(\tt{X } n)^{\tt{rev}}=\tt{X } n$, $(\tt{ctrl}~n~R)^{\tt{rev}}=\tt{ctrl}~n~R^{\tt{rev}}$, $(\tt{swap}~ m~n)^{\tt{rev}}=\tt{swap}~m~n$, $(R_1;~R_2)^{\tt{rev}}=R_2^{\tt{rev}};~ R_1^{\tt{rev}}$.
We prove that the reversed circuit will cancel the behavior of the original circuit.

We can express the semantics of a \rcir program as a function between Boolean registers.
We use notation $[k]_n$ to represent an $n$-bit register storing natural number $k<2^n$ in binary representation. Consecutive registers are labeled sequentially by natural numbers. If $n=1$, we simplify the notation to $[0]$ or $[1]$. 

The translation from \rcir to \sqir is natural since every \rcir construct has a direct correspondence in \sqir. The correctness of this translation states that the behavior of a well-typed classical circuit in \rcir is preserved by the generated quantum circuit in the context of \sqir. That is, the translated quantum circuit turns a state on the computational basis into another one corresponding to the classical state after the execution of the classical reversible circuit.

\paragraph{Details of IMM}
Per \Cref{app:imp:of}, the goal is to construct a reversible circuit $IMM_c(a, N)$ in \rcir satisfying
\begin{align*}
    \forall x<N,~~ [x]_n[0]_s \xrightarrow{IMM_c(a, N)} [a\cdot x \text{ mod } N]_n[0]_s.
\end{align*}
so that we can translate it into a quantum circuit in \sqir.
Adapting the standard practice\cite{ruiz2014algebraic}, we implement modular multiplication based on repeated modular additions.
For addition, we use Cuccaro et al.'s ripple-carry adder (RCA) \cite{cuccaro2004new}. 
RCA realizes the transformation
\begin{align*}
    [c][x]_n[y]_n \xrightarrow{RCA} [c][x]_n[(x+y+c) \text{ mod } 2^n]_n, 
\end{align*}
for ancillary bit $c\in\{0, 1\}$ and inputs $x, y<2^{n-1}.$ 
We use Cucarro et al.'s RCA-based definitions of subtractor (SUB) and comparator (CMP), and we additionally provide a $n$-qubit register swapper (SWP) and shifter (SFT) built using swap gates.
These components realize the following transformations:
\begin{align*}
    [0][x]_n[y]_n &\xrightarrow{SUB} [0][x]_n[(y-x) \text{ mod } 2^n]_n \\
    [0][x]_n[y]_n &\xrightarrow{CMP} [x\geq_{?}y][x]_n[y]_n \\
    [x]_n[y]_n &\xrightarrow{SWP} [y]_n[x]_n \\
    [x]_n &\xrightarrow{SFT} [2x]_n
\end{align*}
Here $x\geq_{?}y=1$ if $x\geq y$, and  $0$ otherwise. SFT is correct only when $x<2^{n-1}$. With these components, we can build a modular adder (\coqe{ModAdd}) and modular shifter (\coqe{ModSft}) using two ancillary bits at positions $0$ and $1$.
\begin{coq}
Definition ModAdd n :=
    $\tt{SWP}_{02}$ n; RCA n; $\tt{SWP}_{02}$ n; CMP n; 
    ctrl 1 (SUB n); $\tt{SWP}_{02}$ n; $\tt{(CMP n)}^{\tt{rev}}$; $\tt{SWP}_{02}$ n.
Definition ModSft n := SFT n; CMP n; ctrl 1 (SUB n).
\end{coq}
$\tt{SWP}_{02}$ is the register swapper applied to the first and third $n$-bit registers. These functions realize the following transformations:
\begin{align*}
    [0][0][N]_n[x]_n[y]_n &\xrightarrow{\texttt{ModAdd}} [0][0][N]_n[x]_n[(x+y) \text{ mod } N]_n \\
    [0][0][N]_n[x]_n &\xrightarrow{\texttt{ModSft}} [0][N\leq_{?}2x][N]_n[2x \text{ mod } N]_n
\end{align*}
Note that $(a\cdot x) \text{ mod } N$ can be decomposed into
\begin{align*}
    (a\cdot x) \text{ mod } N=\left(\sum_{i=0}^{n-1}(1\leq_?a_i)\cdot 2^i\cdot  x\right) \text{ mod } N, 
\end{align*}
where $a_i$ is the $i$-th bit in the little-endian binary representation of $a$. By repeating \coqe{ModSft}s and \coqe{ModAdd}s, we can perform $(a \cdot x) ~\text{mod}~ N$ according to this decomposition, eventually generating a circuit for modular multiplication on two registers ($MM(a, N)$), which implements
\begin{align*}
    [x]_n[0]_n[0]_s \xrightarrow{MM(a, N)} [x]_n[a\cdot  x \text{ mod } N]_n[0]_s.
\end{align*}
Here $s$ is the number of additional ancillary qubits, which is linear to $n$. Finally, to make the operation in-place, we exploit the modular inverse $a^{-1}$ modulo $N$:
\begin{coq}
Definition IMM a N n :=
  MM a N n; $\tt{SWP}_{01}$ n; (MM $\tt{a}^{-1}$ N n$\tt{)}^{\tt{rev}}$.
\end{coq}

There is much space left for optimization in this implementation. Other approaches in the literature \cite{draper2000addition, draper2004logarithmic, van2005fast, pavlidis2012fast, Gidney2021howtofactor} may have a lower depth or fewer ancillary qubits. We chose this approach because its structure is cleaner to express in our language, and its asymptotic complexity is feasible for efficient factorization, which makes it great for mechanized proofs.

\subsection{Implementation of Shor's algorithm}

Our final definition of Shor's algorithm in Coq uses the \coqe{IMM} operation along with a \sqir implementation of \coqe{QPE} described in the previous sections.
The quantum circuit to find the multiplicative order $\text{ord}(a, N)$ is then
\begin{coq}
Definition shor_circuit a N := 
  let m := log2 (2*N^2) in
  let n := log2 (2*N) in
  let f i := IMM (modexp a (2^i) N) N n in
  X (m + n - 1); QPE m f.
\end{coq}
We can extract the distribution of the result of the random procedure of Shor's factorization algorithm
\begin{coq}
Definition factor (a N r : nat) := 
  let cand1 := Nat.gcd (a ^ (r / 2) - 1) N in
  let cand2 := Nat.gcd (a ^ (r / 2) + 1) N in 
  if (1 <? cand1) && (cand1 <? N) then Some cand1
  else if (1 <? cand2) && (cand2 <? N) then Some cand2
  else None.
Definition shor_body N rnd :=
  let m := log2 (2*N^2) in
  let k := 4*log2 (2*N)+11 in
  let distr := join (uniform 1 N) 
                (fun a => run (to_base_ucom (m+k)
                          (shor_circuit a N))) in
  let out := sample distr rnd in
  let a := out / 2^(m+k)) in
  let x := (out mod (2^(m+k))) / 2^k in
  if Nat.gcd a N =? 1
  then factor a N (OF_post a N x n)
  else Some (Nat.gcd a N).
Definition end_to_end_shor N rnds :=
  iterate rnds (shor_body N).
\end{coq}
Here $\tt{factor}$ is the reduction finding non-trivial factors from multiplicative order, $\tt{shor\_body}$ generates the distribution and sampling from it, and $\tt{end\_to\_end\_shor}$ iterates $\tt{shor\_body}$ for multiple times and returns a non-trivial factor if any of them succeeds.

\section{Certification of the Implementation}

In this section, we summarize the facts we have proved in Coq in order to fully verify Shor's algorithm, as presented in the previous section. 

\subsection{Certifying Order Finding}

For the hybrid order finding procedure in \app{imp:of}, we verify that the success probability is at least $1/\text{polylog}(N).$ Recall that the quantum part of order finding uses in-place modular multiplication ($IMM(a, N)$) and quantum phase estimation (QPE). The classical part applies continued fraction expansion to the outcome of quantum measurements. Our statement of order finding correctness says:
\begin{coq}
Lemma Shor_OF_correct :
  forall (a N : nat),
    (1 < a < N) -> (gcd a N = 1) ->
    $\mathbb{P}[$Shor_OF a N = ord a N$]\geq\frac{\beta}{\lfloor \log_2(\texttt{N})\rfloor^4}$.
\end{coq}
where $\beta = \frac{4e^{-2}}{\pi^2}$. The probability sums over possible outputs of the quantum circuit and tests if post-processing finds \coqe{ord a N}.

\paragraph{Certifying IMM}

We have proved that our \rcir implementation of IMM satisfies the equation given in \Cref{sec:imm}.
Therefore, because we have a proved-correct translator from \rcir to \sqir, our \sqir translation of IMM also satisfies this property.
In particular, the in-place modular multiplication circuit $IMM(a, N)$ with $n$ qubits to represent the register and $s$ ancillary qubits, translated from \rcir to \sqir, has the following property for any $0\leq N<2^n$ and $a\in\mathbb{Z}_N$:
\begin{coq}
Definition IMMBehavior a N n s c :=
  forall x : $\mathbb{N}$, x < N ->
    (uc_eval c) $\times ~ (\ket{\tt{x}}_{\tt{n}}\otimes\ket{0}_{\tt{s}})=\ket{\tt{a} \cdot \tt{x mod N}}_{\tt{n}} \otimes \ket{0}_{\tt{s}}$.
Lemma IMM_correct a N :=
  let n := log2 (2*N) in
  let s := 3*n + 11 in
  IMMBehavior a N n s (IMM a n).
\end{coq}
Here \texttt{IMMBehavior} depicts the desired behavior of an in-place modular multiplier, and we have proved the constructed $IMM(a, N)$ satisfies this property.

\paragraph{Certifying QPE over IMM}

We certify that QPE outputs the closest estimate of the eigenvalue's phase corresponding to the input eigenvector with probability no less than $\frac{4}{\pi^2}$:
\begin{coq}
Lemma QPE_semantics :
  forall m n z $\delta$ (f : $\mathbb{N}$ -> base_ucom n) ($\ket{\psi}$ : Vector $2^n$),
    n > 0 -> m > 1 -> $-\frac{1}{2^{\texttt{m}+1}} \leq \delta < \frac{1}{2^{\texttt{m}+1}}$ ->
    Pure_State_Vector $\ket{\psi}$ ->
    (forall k, k < m -> 
      uc_WT (f k) /\ $($uc_eval $($f k$))\ket{\psi}$ = $e^{2^{\texttt{k}+1}\pi i(\frac{\texttt{z}}{2^{\texttt{m}}}+\delta)} \ket{\psi}$) ->
    $\lVert\bra{\texttt{z},\psi} (\texttt{uc\_eval }(\texttt{QPE k n f})) \ket{0, \psi}\rVert^2\geq\frac{4}{\pi^2}$.
\end{coq}

To utilize this lemma with $IMM(a, N)$, we first analyze the eigenpairs of $IMM(a, N)$. Let $r=\text{ord}(a, N)$ be the multiplicative order of $a$ modulo $N$. We define 
\begin{align*}
\ket{\psi_j}_{n}=\frac{1}{\sqrt{r}}\sum_{l<r}\omega_r^{-j\cdot l}|a^l \mod N\rangle_{n}  
\end{align*}
in \sqir~and prove that it is an eigenvector of any circuit satisfying \texttt{IMMBehavior}, including $IMM(a^{2^k}, N)$, with eigenvalue $\omega_r^{j\cdot 2^k}$ for any natural number $k$, where $\omega_r=e^{\frac{2\pi i}{r}}$ is the $r$-th primitive root in the complex plane.
\begin{coq}
Lemma IMMBehavior_eigenpair :
  forall (a r N j n s k : $\mathbb{N}$) (c : base_ucom (n+s)),
    Order a r N -> N < $2^{\tt{n}}$ ->
    IMMBehavior $\tt{a}^{2^\tt{k}}$ N n s c ->
    (uc_eval (f k)) $\ket{\psi_{\tt{j}}}_{\tt{n}} \otimes \ket{0}_{\tt{s}}$ = $e^{2^{\tt{k}+1}\pi i\frac{\tt{j}}{\tt{r}}} \ket{\psi_{\tt{j}}}_{\tt{n}} \otimes \ket{0}_{\tt{s}}.$
\end{coq}
Here \coqe{Order a r N} is a proposition specifying that $r$ is the order of $a$ modulo $N$.
Because we cannot directly prepare $\ket{\psi_j}$, we actually set the eigenvector register in QPE to the state $\ket{1}_{\tt{n}}\otimes\ket{0}_{\tt{s}}$ using the identity:
\begin{coq}
Lemma sum_of_$\psi$_is_one :
  forall a r N n : $\mathbb{N}$,
    Order a r N -> N < $2^{\tt{n}}$ -> $\frac{1}{\sqrt{r}}\sum_{k<r}\ket{\psi_j}_{\tt{n}}=\ket{1}_{\tt{n}}$.
\end{coq}
By applying \coqe{QPE_semantics}, we prove that for any $0\leq k<r,$ with probability no less than $\frac{4}{\pi^2r}$, the result of measuring QPE applied to $\ket{0}_{m}\otimes\ket{1}_{n}\otimes\ket{0}_{s}$ is the closest integer to $\frac{k}{r}2^m.$ 

\paragraph{Certifying Post-processing}

Our certification of post-processing is based on two mathematical results (also formally certified in Coq): the lower bound of Euler's totient function and the Legendre's theorem for continued fraction expansion. 
Let $\mathbb{Z}^*_n$ be the integers smaller than $n$ and coprime to $n$. For a positive integer $n$, Euler's totient function $\varphi(n)$ is the size of $\mathbb{Z}^*_n$. 
They are formulated in Coq as follows. 
\begin{coq}
Theorem Euler_totient_lb : forall n, n $\geq 2$ -> $\frac{\varphi(\tt{n})}{\tt{n}} \geq \frac{e^{-2}}{\lfloor \log_2 \tt{n}\rfloor^4}.$
Lemma Legendre_CFE :
  forall a b p q : $\mathbb{N}$,
    a < b -> gcd p q = 1 -> 0 < q -> $\left|\frac{\tt{a}}{\tt{b}} - \frac{\tt{p}}{\tt{q}}\right| < \frac{1}{2\tt{q}^2}$ ->
    exists s, s $\leq~2\log_2(\tt{b})+1$ /\ CFE s a b = q.
\end{coq}
The verification of these theorems is discussed later.

By Legendre's theorem for CFE, there exists a $s\leq 2m+1$ such that \coqe{CFE s out $2^{\tt{m}}$ = $r$}, where \coqe{out} is the closest integer to $\frac{k}{r}2^m$ for any $k\in\mathbb{Z}^*_r$. Hence the probability of obtaining the order ($r$) is the sum $\sum_{k\in\mathbb{Z}^*_r}\frac{4}{\pi^2 r}$. Note that $r\leq \varphi(N)<N$. With the lower bound on Euler's totient function, we obtain a lower bound of $1/\text{polylog}(N)$ of successfully obtaining the order $r=\text{ord}(a, N)$ through the hybrid algorithm, finishing the proof of \coqe{Shor_OF_correct}.

\paragraph{Lower Bound of Euler's Totient Function}
We build our proof on the formalization of Euler's product formula and Euler's theorem by de Rauglaudre \cite{roglo2020euler}. By rewriting Euler's product formula into exponents, we can scale the formula into exponents of Harmonic sequence $\sum_{0<i\leq n}\frac{1}{i}$. Then an upper bound for the Harmonic sequence suffices for the result. 

In fact, a tighter lower bound of Euler's totient function exists \cite{rosser1962approximate}, but obtaining it involves evolved mathematical techniques which are hard to formalize in Coq since they involved analytic number theory. Fortunately, the formula certified above is sufficient to obtain a success probability of at least $1/\text{polylog}(N)$ for factorizing $N$.

\paragraph{Legendre's Theorem for Continued Fraction Expansion}

The proof of Legendre's theorem consists of facts: (1) \coqe{CFE s a b} monotonically increases, and reaches $b$ within $2\log_2(b)+1$ steps, and (2) for \coqe{CFE s a b $\leq$ q < CFE (s+1) a b} satisfying $\left|\frac{\tt{a}}{\tt{b}} - \frac{\tt{p}}{\tt{q}}\right| < \frac{1}{2\tt{q}^2}$, the only possible value for \tt{q} is \coqe{CFE s a b}. These are certified following basic analysis to the continued fraction expansion\cite{hardy75}.

\subsection{Certifying Shor's Reduction}

We formally certify that for half of the possible choices of $a$, $\tt{ord a n}$ can be used to find a nontrivial factor of $N$:
\begin{coq}
Lemma reduction_fact_OF :
  forall (p k q N : $\mathbb{N}$),
    k > 0 -> prime p -> 2 < p -> 2 < q ->
    gcd p q = 1 -> N = $\tt{p}^\tt{k}*\tt{q}$ ->
    $|\mathbb{Z}_{\tt{N}}| \leq 2 \cdot \sum_{\tt{a}\in\mathbb{Z}_{\tt{N}}} [1 < \tt{gcd (}\tt{a}^{\lfloor\frac{\tt{ord a N}}{2}\rfloor}\pm\tt{1) N < N]}$.
\end{coq}
The expression $[1 < \tt{(gcd (}\tt{a}^{\lfloor\frac{\tt{ord a N}}{2}\rfloor}\pm\tt{1) N) < N]}$ equals to $1$ if at least one of $\texttt{gcd(}\texttt{a}^{\lfloor\frac{\tt{ord a N}}{2}\rfloor}+\texttt{1, N})$ or $\texttt{gcd(}\texttt{a}^{\lfloor\frac{\tt{ord a N}}{2}\rfloor}-\texttt{1, N})$ is a nontrivial factor of $N$, otherwise it equals to 0. In the following we illustrate how we achieve this lemma.

\paragraph{From 2-adic Order to Non-Trivial Factors}

The proof proceeds as follows: Let $d(x)$ be the largest integer $i$ such that $2^i$ is a factor of $x$, which is also known as the 2-adic order. We first certify that $d(\text{ord}(a, p^k))\neq d(\text{ord}(a, q))$ indicates $a^{\lfloor\frac{\text{ord}(a, N)}{2}\rfloor}\not\equiv\pm 1\pmod{N}$
\begin{coq}
Lemma d_neq_sufficient :
  forall a p q N,
    2 < p -> 2 < q -> gcd p q = 1 -> N = pq ->
    d (ord a p) $\neq$ d (ord a q) ->
    $\tt{a}^{\lfloor\frac{\tt{ord a N}}{2}\rfloor}\not\equiv\pm 1\pmod{N}$.
\end{coq}
This condition is sufficient to get a nontrivial factor of $N$ by Euler's theorem and the following lemma
\begin{coq}
Lemma sqr1_not_pm1 :
  forall x N,
    1 < N -> $\tt{x}^2 \equiv 1\pmod{\tt{N}}$ -> $\tt{x} \not\equiv \pm 1\pmod{\tt{N}}$ ->
    1 < gcd (x - 1) N < N \/ 1 < gcd (x + 1) N < N.
\end{coq}

By the Chinese remainder theorem, randomly picking $a$ in $\mathbb{Z}_{N}$ is equivalent to randomly picking $b$ in $\mathbb{Z}_{p^k}$ and randomly picking $c$ in $\mathbb{Z}_{q}$. $a\equiv b\mod{p^k}$ and $a\equiv c\mod{q}$, so $\text{ord}(a, p^k)=\text{ord}(b, p^k)$ and $\text{ord}(a, q)=\text{ord}(c, q).$ Because the random pick of $b$ is independent from the random pick of $c$, it suffices to show that for any integer $i$, at least half of the elements in $\mathbb{Z}_{p^k}$ satisfy $d(\text{ord}(x, p^k))\neq i.$

\paragraph{Detouring to Quadratic Residue}

Shor's original proof\cite{shor1997polynomial} of this property made use of the existence of a group generator of $\mathbb{Z}_{p^k}$, also known as primitive roots, for odd prime $p$. But the existence of primitive roots is non-constructive, hence hard to present in Coq. We manage to detour from primitive roots to quadratic residues in modulus $p^k$ in order to avoid non-constructive proofs.

A quadratic residue modulo $p^k$ is a natural number $a\in\mathbb{Z}_{p^k}$ such that there exists an integer $x$ with $x^2\equiv a\mod{p^k}.$ 
We observe that a quadratic residue $a\in\mathbb{Z}_{p^k}$ will have $d(\text{ord}(x, p^k))<d(\varphi(p^k)),$ where $\varphi$ is the Euler's totient function. Conversely, a quadratic non-residue $a\in\mathbb{Z}_{p^k}$ will have $d(\text{ord}(x, p^k))=d(\varphi(p^k))$:
\begin{coq}
Lemma qr_d_lt :
  forall a p k,
    k $\neq$ 0 -> prime p -> 2 < p -> 
    (exists x, $\tt{x}^2\equiv\tt{a}\mod{\tt{p}^{\tt{k}}}$) ->
    d (ord a $\tt{p}^{\tt{k}}$) < d ($\varphi$ ($\tt{p}^{\tt{k}}$)).
Lemma qnr_d_eq :
  forall a p k,
    k $\neq$ 0 -> prime p -> 2 < p ->
    (forall x, $\tt{x}^2\not\equiv\tt{a}\mod{\tt{p}^{\tt{k}}}$) ->
    d (ord a $\tt{p}^{\tt{k}}$) = d ($\varphi$ ($\tt{p}^{\tt{k}}$)).
\end{coq}
These lemmas are obtained via Euler's Criterion, which describes the difference between multiplicative orders of quadratic residues and quadratic non-residues. The detailed discussion is put later.

We claim that the number of quadratic residues in $\mathbb{Z}_{p^k}$ equals to the number of quadratic non-residues in $\mathbb{Z}_{p^k}$, whose detailed verification is left later. Then no matter what $i$ is, at least half of the elements in $\mathbb{Z}_{p^k}$ satisfy $d(\text{ord}(x, p^k))\neq i$. This makes the probability of finding an $a\in\mathbb{Z}_{p^kq}$ satisfying $d(\text{ord}(a, p^k))\neq d(\text{ord}(a, q))$ at least one half, in which case one of $\tt{gcd }\left(\tt{a}^{\lfloor\frac{\tt{ord a N}}{2}\rfloor}\pm1\right)\tt{ N}$ is a nontrivial factor of $N$.

\paragraph{Euler's Criterion}

We formalize a generalized version of Euler's criterion: for odd prime $p$ and $k>0$, whether an integer $a\in\mathbb{Z}_{p^k}$ is a quadratic residue modulo $p^k$ is determined by the value of $a^{\frac{\varphi(p^k)}{2}} \mod {p^k}$.
\begin{coq}
Lemma Euler_criterion_qr :
  forall a p k,
    k $\neq$ 0 -> prime p -> 2 < p -> gcd a p = 1 ->
    (exists x, $\tt{x}^2 \equiv \tt{a} \mod{\tt{p}^\tt{k}}$) ->
    $\tt{a}^{\frac{\varphi(\tt{p}^\tt{k})}{2}}$ mod $\tt{p}^\tt{k}$ = 1.
Lemma Euler_criterion_qnr :
  forall a p k,
    k $\neq$ 0 -> prime p -> 2 < p -> gcd a p = 1 ->
    (forall x, $\tt{x}^2 \not\equiv \tt{a} \mod{\tt{p}^\tt{k}}$) ->
    $\tt{a}^{\frac{\varphi(\tt{p}^\tt{k})}{2}}$ mod $\tt{p}^\tt{k}$ = $\tt{p}^\tt{k}$ - 1.
\end{coq}
These formulae can be proved by a pairing function over $\mathbb{Z}_{p^k}$:
\begin{align*}
    x \mapsto (a \cdot x^{-1})\mod {p^k},
\end{align*}
where $x^{-1}$ is the multiplicative inverse of $x$ modulo $p^k$. For a quadratic residue $a$, only the two solutions of $x^2\equiv a\mod{p^k}$ do not form pairing: each of them maps to itself. For each pair $(x, y)$ there is $x\cdot y\equiv a\mod{p^k},$ so reordering the product $\prod_{x\in\mathbb{Z}_{p^k}} x$ with this pairing proves the Euler's criterion.

With Euler's criterion, we can reason about the 2-adic order of multiplicative orders for quadratic residues and quadratic non-residues, due to the definition of multiplicative order and $ord(a, p^k)|\varphi(p^k)$.

\paragraph{Counting Quadratic Residues Modulo $p^k$}

For odd prime $p$ and $k>0$, there are exactly $\varphi(p^k)/2$ quadratic residues modulo $p^k$ in $\mathbb{Z}_{p^k}$, and exactly $\varphi(p^k)/2$ quadratic non-residues.
\begin{coq}
Lemma qr_half :
  forall p k,
    k $\neq$ 0 -> prime p -> 2 < p ->
    $|\mathbb{Z}_{\tt{p}^\tt{k}}|=2\cdot\sum_{a\in\mathbb{Z}_{\tt{p}^\tt{k}}}[\exists x, x^2\equiv a\mod{\tt{p}^\tt{k}}].$
Lemma qnr_half :
  forall p k,
    k $\neq$ 0 -> prime p -> 2 < p ->
    $|\mathbb{Z}_{\tt{p}^\tt{k}}|=2\cdot\sum_{a\in\mathbb{Z}_{\tt{p}^\tt{k}}}[\forall x, x^2\not\equiv a\mod{\tt{p}^\tt{k}}].$
\end{coq}
Here $[\exists x, x^2\equiv a\mod{p^k}]$ equals to 1 if $a$ is a quadratic residue modulo $\tt{p}^\tt{k}$, otherwise it equals to 0. Similarly, $[\forall x, x^2\not\equiv a\mod{\tt{p}^\tt{k}}]$ represents whether $a$ is a quadratic non-residue modulo $\tt{p}^\tt{k}$.
These lemmas are proved by the fact that a quadratic residue $a$ has exactly two solutions in $\mathbb{Z}_{p^k}$ to the equation $x^2 \equiv a \mod{p^k}$. Thus for the two-to-one self-map over $\mathbb{Z}_{p^k}$ 
\begin{align*}
    x \mapsto x^2 \text{ mod } p^k,
\end{align*}
the size of its image is exactly half of the size of $\mathbb{Z}_{p^k}$. To prove this result in Coq, we generalize two-to-one functions with mask functions of type $\mathbb{N}\rightarrow\mathbb{B}$ to encode the available positions, then reason by induction.

\subsection{End-to-end Certification}

We present the final statement of the correctness of the end-to-end implementation of Shor's algorithm.
\begin{coq}
Theorem end_to_end_shor_fails_with_low_probability : 
  forall N niter,
    ~ (prime N) -> Odd N -> 
    (forall p k, prime p -> N <> p^k) ->
    $\mathbb{P}_{\tt{rnds}\in \text{Uniform}([0, 1]^{\tt{niter}})}[$end_to_end_shor N rnds = None$]$ 
      <= (1 - (1/2) * ($\beta$ / (log2 N)^4))^niter.
\end{coq}
Then $r$ can be less than an arbitrarily small positive constant $\epsilon$ by enlarging $\tt{niter}$ to $\frac{2}{\beta}\ln\frac{1}{\epsilon}\log_2^4 N$, which is $O(\log^4 N).$

This theorem can be proved by combining the success probability of finding the multiplicative order and the success probability of choosing proper $a$ in the reduction from factorization to order finding. We build an ad-hoc framework for reasoning about discrete probability procedures to express the probability here.

\subsection{Certifying Resource Bounds}

We provide a concrete polynomial upper bound on the resource consumption in our implementation of Shor's algorithm. The aspects of resource consumption considered here are the number of qubits and the number of primitive gates supported by OpenQASM 2.0 \cite{cross2017open}. The number of qubits is easily bounded by the maximal index used in the \sqir program, which is linear to the length of the input. For gate count bounds, we reason about the structure of our circuits. We first generate the gate count bound for the \rcir program, then we transfer this bound to the bound for the \sqir program. Eventually, the resource bound is given by
\begin{coq}
Lemma ugcount_shor_circuit :
  forall a N,
    0 < N ->
    let m := Nat.log2 (2*(N^2)) in
    let n := Nat.log2 (2*N) in
    ugcount (shor_circuit a N) <= 
    (212*n*n + 975*n + 1031)*m + 4*m + m*m.
\end{coq}
Here \texttt{ugcount} counts how many gates are in the circuit. Note $m, n=O(\log N)$. This gives the gate count bound for one iteration as $(212n^2+975n+1031)m+4m+m^2=O(\log^3 N),$ which is asymptotically the same as the original paper \cite{shor1997polynomial}, and similar to other implementations of Shor's algorithm \cite{Gidney2021howtofactor, pavlidis2012fast} (up to $O(\log \log N)$ multiplicative difference because of the different gate sets). 

\section{Running Certified Code}

The codes are certified in Coq, which is a language designed for formal verification. To run the codes realistically and efficiently, extractions to other languages are necessary. Our certification contains the quantum part and the classical part. The quantum part is implemented in \sqir embedded in Coq, and we extract the quantum circuit into OpenQASM 2.0 \cite{cross2017open} format. The classical part is extracted into OCaml code following Coq's extraction mechanism \cite{coqextraction}. Then the OpenQASM codes can be sent to a quantum computer (in our case, a classical simulation of a quantum computer), and OCaml codes are executed on a classical computer. 

With a certification of Shor's algorithm implemented inside Coq, the guarantees of correctness on the extracted codes are strong. However, although our Coq implementation of Shor's algorithm is fully certified, extraction introduces some trusted code outside the scope of our proofs. In particular, we trust that extraction produces OCaml code consistent with our Coq definitions and that we do not introduce errors in our conversion from \sqir to OpenQASM.
We ``tested'' our extraction process by generating order-finding circuits for various sizes and confirming that they produce the expected results in a simulator.

\subsection{Extraction}

For the quantum part, we extract the Coq program generating \sqir circuits into the OCaml program generating the corresponding OpenQASM 2.0 assembly file. We substitute the OpenQASM 2.0 gate set for the basic gate set in \sqir, which is extended with: $X, H, U_1, U_2, U_3, CU_1, SWAP, CSWAP, CX, CCX, C3X, C4X$. Here $X, H$ are the Pauli $X$ gate and Hadamard gate. $U_1, U_2, U_3$ are single-qubit rotation gates with different parametrization \cite{cross2017open}. $CU_1$ is the controlled version of the $U_1$ gate. $SWAP$ and $CSWAP$ are the swap gate and its controlled version. $CX, CCX, C3X$, and $C4X$ are the controlled versions of the $X$ gate, with a different number of control qubits. Specifically, $CX$ is the CNOT gate. The proofs are adapted with this gate set. The translation from \sqir to OpenQASM then is direct. 

For the classical part, we follow Coq's extraction mechanism. We extract the integer types in Coq's proof to OCaml's \texttt{Z} type, and several number theory functions to their correspondence in OCaml with the same behavior but better efficiency. Since our proofs are for programs with classical probabilistic procedures and quantum procedures, we extract the sampling procedures with OCaml's built-in randomization library.

One potential gap in our extraction of Coq to OCaml is the assumption that OCaml floats satisfy the same properties as Coq Real numbers. It is actually not the case, but we did not observe any error  introduced by this assumption in our testing. In our development, we use Coq's axiomatized representation of reals \cite{coqreals}, which cannot be directly extracted to OCaml. We chose to extract it to the most similar native data type in OCaml--floating-point numbers. 
An alternative would be to prove Shor's algorithm correct with gate parameters represented using some Coq formalism for floating-point numbers \cite{floqc},  which we leave for future work.

\subsection{Experiments}

We test the extracted codes by running small examples on them. Since nowadays quantum computers are still not capable of running quantum circuits as large as generated Shor's factorization circuits ($\sim$30 qubits, $\sim 10^4$ gates for small cases), we run the circuits with the DDSIM simulator \cite{ddsim} on a laptop with an Intel Core i7-8705G CPU.  The experiment results are included in \Cref{fig:experiments} (b) (c).

As a simple illustration, we showcase the order finding for $a=3$ and $N=7$ on the left of \Cref{fig:experiments} (b). The extracted OpenQASM file makes use of 29 qubits and contains around 11000 gates. DDSIM simulator executes the file and generates simulated outcomes for $10^5$ shots. The measurement results of QPE are interpreted in binary representation as estimated $2^m\cdot k/r$. In this case, the outcome ranges from 0 to 63, with different frequencies. We apply OCaml post-processing codes for order finding on each outcome to find the order. Those measurement outcomes reporting the correct order (which is 6) are marked green in \Cref{fig:experiments} (b). The frequency summation of these measurement outcomes over the total is 28.40\%, above the proven lower bound of the success probability of order finding which is 0.17\% for this input. 

We are also able to simulate the factorization algorithm for $N=15$. For any $a$ coprime to 15, the extracted OpenQASM codes contain around 35 qubits and 22000 gates. Fortunately, DDSIM still works efficiently on these cases due to the well-structured states of these cases, taking around 10 seconds for each simulation. 
We take $7\times 10^5$ shots in total. When $N=15$, the measurement outcomes from QPE in order finding are limited to $0, 64, 128, 192$ because the order of any $a$ coprime to 15 is either $2$ or $4$, so $2^m\cdot k/r$ can be precisely expressed as one of them without approximation. 
The frequency of the simulation outcomes for $N=15$ is displayed on the right of \Cref{fig:experiments} (b). We then apply the extracted OCaml post-processing codes for factorization to obtain a non-trivial factor of $N$. The overall empirical success probability is 43.77\%, above our certified lower bound of 0.17\%.

We have also tested larger cases on DDSIM simulator \cite{ddsim} for input size ranging from $2$ bits to $10$ bits (correspondingly, $N$ from $3$ to $1023$), as in \Cref{fig:experiments} (c). Since the circuits generated are large, most of the circuits cannot be simulated in a reasonable amount of time (we set the termination threshold 1 hour). We exhibit selected cases that DDSIM is capable of simulating: $N=15, 21, 51, 55, 63, 77, 105, 255$ for factorization, and $(a, N)=(2, 3), (3, 7), (7, 15), (4, 21), (18, 41), (39, 61), (99, 170)$, $(101, 384), (97, 1020)$ for order finding. These empirically investigated cases are drawn as red circles in \Cref{fig:experiments} (c). Most larger circuits that are simulated by DDSIM have the multiplicative order a power of $2$ so that the simulated state is efficiently expressible. 
For each input size, we also calculate the success probability for each possible input combination by using the analytical formulae of the success probability with concrete inputs. Shor shows the probability of obtaining a specific output for order finding is\cite{shor1997polynomial} 
\begin{align*}
    \mathbb{P}[\texttt{out}=u]=\frac{1}{2^{2m}}\sum_{0\leq k<r}\left\lvert\sum_{\substack{0\leq v<r \\ v\equiv k\pmod{r}}} e^{2\pi iuv/2^m}\right\rvert^2.
\end{align*}
Here $r$ is the order, and $m$ is the precision used in QPE. The success probability of order finding then is a summation of those $u$ for which the post-processing gives correct $r$. For most output $u$, the probability is negligible. The output tends to be around $2^m k/r$, so the sum is taken over integers whose distance to the closest $2^m k/r$ (for some $k$) is less than a threshold, and the overall probability of getting these integers is at least 95\%. Hence the additive error is less than 0.05.
These empirical results are drawn as blue intervals (i.e., minimal to maximal success probability) in \Cref{fig:experiments} for each input size, which is called the empirical range of success probability. 
The certified probability lower bounds are drawn as red curves in \Cref{fig:experiments} as well. 
The empirical bounds are significantly larger than the certified bounds for small input sizes because of loose scaling in proofs, and non-optimality in our certification of Euler's totient function's lower bounds. 
Nevertheless, asymptotically our certified lower bound is sufficient for showing that Shor's algorithm succeeds in polynomial time with large probability. 

We also exhibit the empirical gate count and certified gate count for order finding and factorization circuits. In fact, the circuits for order finding are exactly the factorization circuits after $a$ is picked, so we do not distinguish these two problems for gate count.
On the right of \Cref{fig:experiments} (c), we exhibit these data for input sizes ranging from 2 to 10. We enumerate all the inputs for these cases and calculate the maximal, minimal, and average gate count and draw them as blue curves and intervals. The certified gate count only depends on the input size, which is drawn in red. One can see the empirical results satisfy the certified bounds on gate count. Due to some scaling factors in the analytical gate count analysis, the certified bounds are relatively loose. Asymptotically, our certified gate count is the same as the original paper's analysis.

\end{document}